# Asymmetric Transmission in Photonic Structures with Phase-Change Components

N. Antonellis[1], R. Thomas[1], M. A. Kats[2], I. Vitebskiy[3] and T. Kottos[1]

[1]*Department of Physics, Wesleyan University, Middletown, 06459, United States*

[2]*Department of Electrical and Computer Engineering, University of Wisconsin – Madison, Madison 53706, United States*

[3]*Air Force Research Laboratory, Sensors Directorate, Wright-Patterson Air Force Base, OH-45433, United States*

*Abstract-* We consider the scattering problem for an asymmetric composite photonic structure with a component experiencing a thermally driven phase transition. Using a numerical example, we show that if the heating is caused by the incident light, the transmittance can become highly asymmetric within a broad range of light intensities. This effect can be utilized for directional light transmission, asymmetric optical limiting, or power switching.

*1. Introduction-* Control of the directionality of electromagnetic radiation is a recurring theme of research for both physicists and engineers. On the fundamental level, this challenge is directly related to the restrictions imposed by the reciprocity principle, which applies to light propagation in media without spontaneous magnetic order and in the absence of a magnetic-field bias [1]. Although the magneto-optical approach to asymmetric or unidirectional light propagation has been dominant in optics and the microwave regime [2-7], it has some inherent problems, which have stimulated great interest in alternative ways to break the reciprocity principle and to achieve strong asymmetry between forward and backward light transmission. One popular non-magnetic approach is to use temporal modulation, which is an active scheme to break reciprocity [8-13]. In this study, we restrict ourselves to passive techniques.

Passive non-magnetic approaches to asymmetric light propagation usually rely on nonlinear spatially asymmetric structures. The nonlinear effects can be different for the forward (*F*) and backward (*B*) propagating light, thus resulting in intensity-dependent propagation asymmetry [14-21]. In all cases, at low input light intensity, the transmittance becomes symmetric due to the reciprocity principle.

Another non-magnetic approach to transmission asymmetry is based on thermal effects produced by incident light in spatially asymmetric structures involving components with temperature-dependent material (optical) parameters [22-24]. Specifically, input light of the same intensity incident from opposite sides of an asymmetric structure can produce different heating effects and, therefore, the transmission coefficient for forward and backward propagation can also be different. The advantages of the approach based on light-induced heating over using asymmetric structures with conventional nonlinear materials can include: (i) a much lower input light intensity required to trigger a change in the temperature-dependent material parameters, and (ii) a change in the complex index being much larger than that of conventional nonlinear effects [25]. The light-induced asymmetric heating is dependent on thermal conductivity and other heat transfer mechanisms [26,27]. This provides additional flexibility in design but, on the other hand, the thermal transport significantly complicates the theoretical modeling and the associated numerical simulations.

The most effective manifestations of heating-related effects in light propagation occur if the optical material exhibits a nearly abrupt, heat-induced phase transition, rather than a gradual change in the refractive index and/or absorption. A well-known example of such phase-change materials (PCMs) is vanadium dioxide ($VO_2$) which undergoes an insulator-to-metal phase change when heated above approximately $\theta_c = 68 \,^\circ C$ [28-30]. Asymmetric light transmission associated with heat-induced phase change in the $VO_2$ component of a trilayer gold/$VO_2$/sapphire structure was successfully demonstrated in Ref. [31]. One limitation of such a simple three-layer asymmetric structure is that at optical frequencies, the heat-induced change in the complex refractive indices of all known PCMs, including $VO_2$, may not be sufficient to produce a change in the transmission characteristics of the stack that is significant enough for the structure to act as an effective limiter, switch, or an isolator. Furthermore, a few-layer structure provides limited ability to control the value of the input light intensity that would trigger the phase transition. The ability to control that critical value is very important, since the desired values of the limiting/switching threshold can vary significantly in different applications.

To address the above limitations, we propose to incorporate a PCM into an asymmetric multi-layered photonic structure (MLPS). In the simplest realization, a quarter-wavelength-thick PCM layer is used as a defect layer which supports a localized mode within a periodic MLPS, as shown in Fig. 1(a). In other words, we consider an asymmetric optical microcavity filled with a PCM. The asymmetry of the microcavity is essential for light-induced transmission asymmetry. Using time-domain simulations for Maxwell's equations coupled with heat-transfer equations, we show that for a range of incident light fluences, the PCM-filled microcavity displays a highly asymmetric transmittance. Specifically, the asymmetric photonic structure in Fig. 1 is highly transmissive in the forward direction at the microcavity resonance frequency within a broad range of the input light intensity. By contrast, in the backward direction, the same structure is highly reflective for the same input light intensity and within the frequency range covering the entire photonic band gap. For a given PCM layer ($VO_2$ in our case), the lower and the upper limits of the incident light intensity within which the structure displays unidirectional transmittance, can be engineered by the proper choice of the layered structure geometry. The frequency of the unidirectional (forward) resonant transmittance can also be changed significantly.

*2. Underlying basic principle of asymmetric transport* – Before analyzing the asymmetric transport within the MLPS of Fig. 1, we demonstrate the underlying principle by studying the transport properties of a toy model. We first consider the case of incoming waves with low incident power and the consequences of spatial asymmetry in the optical potential defining a cavity. A simple model consisting of three delta-function-like layers at positions $z_1 = 0$, $z_3 = 1$ (total length of the structure; which will be considered as our unit of length), and at $0 \leq z_D \leq 1$ can effectively demonstrate the dependence of transport characteristics of such cavity as a function of the optical asymmetry. While the two delta-like layers at $z_1$, and $z_3$ define the extend of the cavity, the position $z_D$ of the third (defect) layer can control the degree of the optical asymmetry. The Helmholtz equation that describes the transport properties of a monochromatic wave through this cavity is

$$\left(\frac{d^2}{dz^2} + k^2\left[n^2 \sum_{l=1}^{3} \delta(z - z_l) + n_0^2\right]\right) E = 0, \tag{1}$$

where $z$ is the position (in units of $z_3$) along the propagation distance, $E(z)$ is the electric field amplitude at position $z$, $n$ is the refractive index of the thin layers, $n_0 = 1$ is the refractive index of air, $k = \frac{2\pi}{\lambda_0}$ is the wave number of the incident wave and $\lambda_0$ is the wavelength in free space. For a greater clarity, we assumed that the index of refraction $n$ of all delta-like layers is purely real and the same (this can be justified when the incident power is low -- in this case, the defect layer at $z_D$ is assumed to be in the insulating phase where for a $VO_2$ material one can assume that $\varepsilon_D'' = \varepsilon_{VO_2}'' \approx 0$ (see Fig. 1b) ). The Helmholtz equation (1) can be easily solved analytically under appropriate scattering boundary conditions: for incident waves in the forward direction, we assume the electric field amplitude is:

$$E(z < z_1) = e^{-ikz} + r_F e^{ikz} \text{ and } E(z > z_3) = t_F e^{-ikz}. \tag{2}$$

Similarly, for incident waves in the backward direction we shall have:

$$E(z > z_3) = e^{ikz} + r_B e^{-ikz} \text{ and } E(z < z_1) = t_B e^{ikz} \tag{3}$$

where $r_{F/B}$ and $t_{F/B}$ denote the amplitudes of the reflected and transmitted waves, respectively, for the incident waves in the forward/backward directions. Imposing continuity of the field and allowing the discontinuity of its first derivative at positions $z_1$ $z_D$, and $z_3$, we can obtain the spatial electric field distribution $E(z)$. Figures 2(a)-(b) show two scattering electric-field intensity profiles denoted by black (red) for forward (backward) incident waves with the same incident field amplitude, for two different positions of the middle delta-like layer: (a) $z_D = 0.5$ where mirror symmetry is preserved and (b) $z_D \neq 0.5$ where mirror symmetry is violated. In the former case, the two field intensities are mirror images of one another; *i.e.*, a transformation $z \leftrightarrow -z$ maps one solution to the other. In contrast, Fig. 2b shows that the left and right field intensities are completely different from one another. The field intensity difference $\Delta(z_D)$, between the scattering fields associated with a forward and backward propagating incident waves is given by:

$$\Delta(z_D) \equiv |E_F(z_D)|^2 - |E_B(z_D)|^2 = 16kn\big(-2\cos(k) + kn\sin(k)\big)\sin(-k + 2kz_D)B(k,n,z_D), \quad (4)$$

where all lengths (wavenumbers) are measured in units of the (inverse) length of the cavity $z_3 (= 1)$ and

$$B^{-1}(k,n,z_D) = 0.25 | e^{2ik(1+z_D)}k^2n^2(kn - 2i) + (kn + 2i)(k^2n^2(e^{2ik} + e^{4ikz_D}) - e^{2ikz_D}(2i + kn)^2)|^2. \quad (5)$$

The forward and backward transmittances are, nevertheless, the same; i.e., $T_F \equiv |t_F|^2 = |t_B|^2 \equiv T_B$ due to the Lorentz reciprocity theorem.

The forward/backward E-field intensity asymmetry $\Delta(z_D)$ can lead to directional transport when the defect at position $z_D$ consists of a material that has intensity-dependent optical parameters. Let us, for example, consider that the defect layer consists of a PCM that shows a strong and abrupt change in its complex permittivity due to an insulator-to-metal transition when its temperature $\theta_D$ exceeds a critical value $\theta_C$. These abrupt temperature-driven permittivity variations are in contrast to gradual Kerr-type or to two photon absorption nonlinearities, and can be triggered by heating of the defect layer due to absorbed incident radiation.

A simple calculation (see for example [33]) allow us to connect the absorbance $A_D$ at the defect layer with the associated field intensity i.e. $A_D \sim \Im m(\varepsilon_D)\omega |E(z_D)|^2$. Specifically, the PCM layer experiences different absorbance, with difference $\Delta A = |A_F - A_B| \sim \Im m(\varepsilon_D)\omega|\Delta(z_D)| \neq 0$, for forward/backward propagating incident waves with the same incident irradiances. Consequently, for forward/backward incident waves, the PCM will develop different temperatures $\Delta\theta \equiv |\theta_F - \theta_B|$ which will affect differently the imaginary value of its permittivity; i.e., $\varepsilon_D''(\theta_F) \neq \varepsilon_D''(\theta_B)$. Note that there is also a change in real permittivity of the PCM with temperature, but it is not as dramatic as the change in the imaginary component (see Fig. 1(b)). The most dramatic difference appears when the intensity of the (say backwards) scattering field, at the position of the defect layer $z_D$ is such that $A_B > A_F$, leading to a temperature $\theta_B = \theta_C > \theta_F$. In this case, $\varepsilon_D''(\theta_F) \ll \varepsilon_D''(\theta_B)$, and thus the forward and backward transmittances will be dramatically different. Specifically, a sudden jump in the $\varepsilon_D''(\theta_B)$ by three-four orders of magnitude (insulator to metallic phase) is followed by an under-damping to an over-damping transition resulting to a suppression of the transmittance and of absorbance (as opposed to the reflectance which can reach values close to unity). In section 5 we will further explain the origin of this transition.

In Figs. 2c,d we plot an example of the spatial electric-field intensity distributions for the case of a stand-alone (SA) structure of $VO_2$ on a ZnS substrate. It is important to point out that the field asymmetry in the cases of Figs. 2c,d and in the case of our delta-like layer model (see Fig. 2b) is moderate (*i.e.*, two or three-fold) and thus any directional asymmetry in insulator-metal transition (IMT) happens for a small range of incident electric field intensities, see solid symbols in Fig. 1c. Moreover, the values of the field intensity at the position of the $VO_2$ layer are comparable with the incident field intensities. Therefore, in order to heat up the PCM up to $\theta_C$ and trigger the IMT, the incident irradiance must be relatively high.

*3. Multi-Layered Photonic Structure* – The multi-layered photonic structure (MLPS) that we study in this paper consists of alternating layers of high-index (*H*) and low-index (*L*) quarter-wavelength-thick dielectric films which produces transmission and reflection bands. Inside the MLPS we place a defect layer *D* at a position away from the mirror-symmetry plane of the structure, see for example Fig. 1(a). The proposed asymmetric MLPS is represented by the form *(LH)$^m$[D](LH)$^n$* where $m \neq n$ corresponds to $m(n)$ bilayers on the left (right)-hand side of the defect layer. Furthermore, we assume that the defect layer is also quarter-wavelength thick and is made of a phase-change material (PCM) that undergoes an IMT when the temperature of the layer reaches a critical value, $\theta_C$.

The MLPS has been designed such that it supports a resonant defect mode (due to the presence of the defect layer) which is exponentially localized around the PCM defect layer and is located at the center of the band gap at $\lambda_0 \sim 10.5 \, \mu m$. We note that this wavelength corresponds to the middle of the long-wave infrared atmospheric transparency window. In our numerical example below, we have chosen (crystalline) silicon with index of refraction $n_{Si} = 3.4$ and Zinc Sulfide (ZnS) with index of refraction $n_{ZnS} = 2.2$ as the high- and low-index dielectric materials. We have checked that a quantitatively similar transport behavior can be obtained also in case that the low-index material is chosen to be Zinc Selenide (*ZnSe*) with $n_{ZnSe} = 2.4$ and/or the high index material is (amorphous) silicon with refractive index $n_{Si} = 3.7$ In all these cases, we have assumed that these materials are lossless around $10.5 \, \mu m$ wavelength [34-35].

The VO₂ PCM layer undergoes an IMT at $\theta_C \approx 68$ °C which changes both its real and imaginary permittivity as a function of temperature, $\theta$. In our simulations, the complex permittivity $\varepsilon(\theta) = \varepsilon'(\theta) - i\varepsilon''(\theta)$ of the VO₂ defect layer was obtained directly from the experimentally reported values for heating (solid symbols), measured at $\lambda_0 = 10.5$ $\mu m$ (see Fig.1b) [32]. For simulation purposes, the values of $\varepsilon'$ and $\varepsilon''$ from 20 to 72 °C were linearly interpolated, and outside this range we assumed them to be constant.

The latent heat released during the IMT of the VO₂ layer leads to a temperature dependence of the specific heat capacity, $C_p^D$, which can be approximately related with the slope of the change in electrical conductivity with temperature: $C_p^D(\theta) = C_p^0 + \frac{H_L}{\Delta\sigma_t}\frac{d\sigma}{d\theta}$ [36]. $C_p^0 = 690$ $\frac{J}{kgK}$ corresponds to the specific heat capacity of VO₂ before the phase transition and we have assumed the latent heat to be constant, $H_L \approx 5.04 \times 10^4$ $J/kg$ [37,38]. Finally, $\Delta\sigma_t = 18.64 \cdot 10^4 (S/m)$ is the total conductivity jump during the phase transition extracted from the data of $\varepsilon''(\theta)$ of VO2 (see Fig. 1b).

*4. Transient and Steady-State Modeling* --The wave propagation along the *z*-direction is described by:

$$\nabla \times \vec{H} = \vec{j}_0 + \varepsilon(z)\frac{\partial \vec{E}}{\partial t}, \quad \nabla \times \vec{E} = -\mu_0 \frac{\partial \vec{H}}{\partial t}, \tag{6}$$

where $\varepsilon(z) = \varepsilon'(z) - i\varepsilon''(z)$ is the position-dependent permittivity that varies from layer to layer, $\mu_0$ is the free space permeability, and $\vec{j}_0 = \sigma_0(z)\vec{E}$ is the electric current density. Equations (6) have been solved together with the transient heat-transfer rate equations, Eq. (7), which give the temperature variation of the PCM layer in the presence of continuous wave (CW) incident radiation (see Fig. 1b):

$$\rho_D C_p^D(\theta)\frac{\partial\theta}{\partial t} - \nabla \cdot (k_D \nabla\theta) = Q(\theta) + q_0 + q_r, \tag{7}$$

where $\rho_D = 4670$ kg/$m^3$ and $k_D = 4$ W/m·K denote the mass density, and thermal conductivity, of the defect layer [39,40]. The heat production $Q$ per unit volume of the PCM layer is $Q = \frac{1}{2}[Re(\vec{j}_0 \cdot \vec{E}) + \omega\varepsilon''(z)Re(\vec{E} \cdot \vec{E})]$. For computational convenience, we recast the above expression as $Q = \frac{1}{2}Re(\vec{j} \cdot \vec{E})$, where $\vec{j} = \sigma\vec{E}$, $\sigma = \sigma_0 + \omega\varepsilon''$. In our simulations of electromagnetic transport we have ignored dispersion phenomena of $\varepsilon(z)$. Indeed, the change in the permittivity of the VO₂ layer due to the IMT (via a self-induced heating) which is responsible for the asymmetric limiting action is three-four orders of magnitude, which is much greater compared to any variations due to dispersion effects.

The term $q_0 = h(\theta_{ext} - \theta)$ denotes the thermal convection occurring from the edges of the dielectric layer to the surrounding air, and $h = 10$W/$m^2$ K and $\theta_{ext} = 293.15$ K corresponds to heat flux coefficient and temperature of the air [41]. Finally, $q_r = \epsilon_r \sigma_r(\theta_{ext}^4 - \theta^4)$ describes, the heat transfer via thermal radiation from the edges of the photonic structure to the surrounding air assuming grey-body approximation for the Si and ZnS layers [42]. The parameters $\epsilon_r$ and $\sigma_r$ (= 5.7x $10^{-8}$ $W \cdot m^{-2}K^{-4}$) correspond to the thermal emissivity coefficient of the edge layers and the Stefan-Boltzmann constant, respectively. In our simulations, the thermal emissivity from each of the two edge layers (Si or ZnS) of the structure is estimated using Kirchoff's law $\epsilon_r^{L/R} = 1 - T^{L/R} - R^{L/R}$ [42]. The transmittance $T^{L/R}$ and reflectance $R^{L/R}$, associated with a left/right (L/R) incident wave respectively, have been evaluated using Maxwell's equations (6) (without coupling them with the thermal transport equations) for various values of the permittivity (corresponding to different temperatures) of the PCM (see Fig. 1b). Then, the calculated $\epsilon_r^{L/R}(\varepsilon_D)$ has been used in the coupled Maxwell's-heat transfer equations. The consistency of the scheme has been checked by comparing the used value of $\epsilon_r^{L/R}$ with the asymptotic absorbance calculated using the coupled Eqs. (6,7) (see Fig. 3e).

Equations (6,7) were solved self-consistently using a finite-element software package from COMSOL MULTIPHYSICS [43] to numerically evaluate the temporal behavior and steady-state values of transmittance reflectance, absorbance, and of the temperature of the VO2 defect layer. For the transient analysis we have used a Frequency-Transient Electromagnetic module coupled to heat transfer equations which allowed us to numerically evaluate the temporal behavior of transmittance $T(t)$, reflectance $R(t)$ and absorbance $A(t)$= 1- $T(t)$- $R(t)$ of the asymmetric MLPS, and the temperature $\theta(t)$ at the PCM defect layer. From these calculations, we extracted the asymptotic (steady-state) values as t→ ∞ of these quantities $T_\infty, R_\infty, A_\infty$ and $\theta_\infty$ as well as the steady-state electric field profiles. The extracted steady state values of transmittance, reflectance and absorbance and defect temperature i.e $T_\infty, R_\infty, A_\infty$ and $\theta_\infty$ respectively, have been also compared with the results calculated using a Frequency-Stationary module coupled to heat transfer equations. In the simulations we used a varying mesh: the silicon and ZnSe layers have been partitioned with 50-150 elements per layer, depending on the simulation, while the $VO_2$ layer has been partitioned with 300-1000 elements. The convergence of the results has been evaluated with a tolerance factor which has been set to .1%. We have furthermore repeated the calculations by doubling the number of mesh points in order to guarantee the accuracy of the converged numerical solutions.

*5. Results and Analysis-* Equations (6) and (7) were solved for both cases when light enters the MLPS in the forward ($F$) and backward ($B$) directions, respectively. Figures 2e,f show the steady-state electric-field intensity profiles of a MLPS with $m = 10$ and $n = 4$ bilayers at the resonant frequency ($\nu_0 = 2.86 \times 10^{13}$ Hz) corresponding to irradiances of $I$=2.66x10$^{-5}$ W/cm$^2$ (blue lines), $I$=1.06x10$^{-2}$ W/cm$^2$ (black lines) and $I$=0.39 W/cm$^2$ (dashed red lines).

We utilized the exponential shape of the resonant defect mode (localized around the position of the defect) in order to achieve $\theta_C$ inside the PCM (and thus induce the IMT), for values of incident field irradiances that are exponentially lower in comparison to the single stand-alone structure (see Fig. 1c). Another consequence of the MLPS is the appearance of an enhanced forward-backward electric-field intensity asymmetry $\Delta(z_2)$. For low irradiances (solid blue and black lines), resulting in $\theta_F, \theta_B < \theta_C$, the MLPS demonstrates high transmittance ($T_\infty \approx 0.6$) at the defect-mode frequency, see Fig. 3(a), which is the same for left and right incident waves. This is associated with resonant transport via the defect mode. At the same time, the field intensity at the PCM layer is higher for waves incident in the backward direction (Fig. 2f) than for waves incident in the forward direction (Fig. 2e). This leads us to the conclusion that an incident wave in the backward direction will heat up the PCM to its $\theta_C$ at smaller irradiances than its forward propagating counterpart. Indeed, for higher irradiances (dashed red lines), light entering the structure in the forward direction (i.e., from left) does not induce critical heating and thus the PCM remains in the dielectric phase (Fig. 1c). In this case, the presence of the resonant defect mode (dashed red line in Fig. 2e) leads to high transmission, similar to the case of low irradiances (solid black and blue lines). When the wave with the same high irradiance enters the MLPS in the backward direction (i.e., from the right), it causes heating of the $VO_2$ layer above $\theta_C$ (see Fig. 1b) which drives the PCM to the metallic phase. The dramatic increase in $\varepsilon_D''(\theta)$ greatly reduces the quality factor of the resonant mode and increases the impendence mismatch of the resonant mode with the incoming wave. As a result, the reflection increases dramatically with a simultaneous decrease in both the transmission and absorption.

The increased reflectivity can be further understood in terms of an under-damping to over-damping transition (as discussed at the end of section 2). The latter is controlled by the interplay between the radiative losses, occurring at the edge of the photonic structures, and the bulk Ohmic losses associated with the $VO_2$ defect layer. The former is exponentially small due to the (exponential) shape of the resonant defect mode (see Figs. 2ef) which is localized around the defect layer. On the other hand, the strength of the bulk losses depends on the the value of the imaginary part of the complex permittivity of the $VO_2$ layer which is controlled by the value of the field intensity at the position of the $VO_2$ defect (see discussion in previous paragraph and at section 2). As $\varepsilon_D''$ increases due to heating (induced as the incident radiation becomes high enough), the bulk losses overrun the radiative losses and spoil the resonance "driving" the system to the over-damping regime. Consequently, the incident CW signal cannot coupled to the (spoiled) resonant mode and instead it is reflected back in space. We stress again that the dependence of the value of the field profile at the position of the defect from the direction of the incident wave, is the main source of the right/left asymmetric IMT experienced by the $VO_2$. This asymmetry leads to an earlier/later triggering of the under-damping to over-damping transition and it is the source of the asymmetric transport.

The qualitative difference in the transport of the MLPS for forward and backward incident waves occurs for a broad range of incident electric field irradiances—as compared to the case of a single bilayer structure—and will be confirmed via detailed multi-physics simulations in the rest of the paper.

Figures 3a-c display some typical transient behaviors of the transmittance, $T(t)$, for a CW incident signal propagating in the forward (black lines) and backward (red lines) directions with small (a), moderate (b) and large irradiances (c), respectively. When the irradiance of the CW signal is small ($I = 0.033$ W/cm$^2$), the heating of the VO$_2$ defect layer is negligible for light incident in both directions. As a result, the PCM remains in the dielectric phase irrespective of the direction of the incident light. Thus, the resonant defect mode is unaffected, and the photonic structure remains (almost) transparent at the resonant frequency, as indicated in Fig. 3(a). When the input irradiance is increased, the strong asymmetry in the distribution of the electric field intensity kicks in, leading to an asymmetric heating and consequent IMT of the VO$_2$ defect layer. We found that for a range of irradiances, $0.08 < I < 0.4$ W/cm$^2$, the forward propagating incident wave does not cause any significant heating of the VO$_2$ defect layer, and thus the photonic structure remains transparent at all times (black curve in Fig. 3(b)). In contrast, for the same range of irradiances, a backward propagating incident wave causes critical heating of the PCM layer at a much earlier time, which triggers the IMT and a corresponding jump in the value of $\varepsilon_D''(\theta)$ (see Fig. 1(c)). This, in turn, suppresses the resonant mode and creates an impedance mismatch with the incident CW wave. Consequently, the transmittance, $T_B$, of the backward propagating incident radiation drops abruptly by almost three orders in magnitude (see red curve in Fig. 3(b)). Further increase of the irradiance leads to the triggering of the IMT for forward propagating incident waves as well. This IMT occurs, though, at later times (i.e., at $t_F \approx 0.3$ sec) with respect to the one associated with backward propagating incident radiation (i.e., at $t_B \approx 0.03$ sec). In this case, the MLPS acts as a double-sided reflector, see Fig. 3(c). Note that both time scales can be exponentially scaled-down by increasing the number of bilayers on the left and right side of the PCM [27].

An overview of the directional behaviour of the steady state transmittance $T_\infty$ versus irradiance is shown in Fig. 3(d) for the case of the MLPS (continuous lines) and for the case of an SA VO$_2$-ZnS bilayer (dashed lines). In this figure, the forward ($T_F$) and backward ($T_B$) transmittances are shown with black and red lines. The results are obtained by solving Eqs. (1,2) in the limit when t→ ∞ (steady state). The drop in transmittance occurs for irradiances that are more than an order of magnitude smaller than that of a stand-alone structure. At the same time, the range of irradiances for which asymmetric transmission occurs is dramatically increased for the MLPS (as compared to the single bilayer).

In Fig. 3e we also report the steady state absorbance, $A_\infty$. For a forward-propagating incident wave, the absorbance initially increases and then decreases (red open circles), following the typical scenario (see discussion at the end of section 2) associated with an underdamping (domain of increasing absorbance) to over-damping transition (domain of decreasing absorbance). The initial increase of $A_\infty$ is associated with the increase of $\varepsilon_D''(\theta)$, while its subsequent decrease is associated with the suppression of the resonant defect mode and the high reflectivity induced by the impedance mismatch. This is in contrast to the SA layered structure, where an absorption as high as $A_\infty = 0.6$ is obtained, which can lead to overheating of the PCM layer, and its destruction. Therefore, the asymmetric MLPS can act also as a highly asymmetric limiter [26, 33,44-46].

In order to guarantee the stability of the extracted $T_\infty$, $A_\infty$ values we repeated the simulations by solving Eqs. (1,2) under steady state conditions, using a Frequency-Stationary modulo of COMSOL MULTIPHYSICS [43]. Furthermore, we have checked that the reported values are associated with the maxima values of $T_\infty$, $A_\infty$ within the band-gap by calculating them for a number of resonant frequencies inside the band-gap. In this way we were certain that the total transmittance is indeed destroyed and the reported drop in transmittance is not a consequence of a resonance shift.

Our simulations directly apply to the case when: (a) both the lateral dimensions of the layered structure and the beam diameter are much greater than the stack thickness, which is approximately 0.03 mm and (b) the light intensity is uniform within the beam cross-section. The assumption (a) is quite realistic when it comes to a wide-aperture design for a free-space setting. If, on the other hand, the condition (a) is satisfied, but the beam profile is not uniform, then the light-induced heating will also be dependent on the lateral coordinates *x* and *y*. Assuming that the beam profile is smooth enough, we can still apply our simulations to this case, but the temperature and the stack (local) transmittance will be dependent on the lateral coordinates. The latter feature is common for all nonlinear or heat-related effects caused by pulses with non-uniform profile, but in our photonic design, the transmittance non-uniformity will be enhanced by the resonant conditions. In practice, the non-uniformity of the pulse profile will smooth out the otherwise abrupt transition from high transmittance to high reflectivity, as the light intensity or fluence exceed the limiting threshold.

*6. Conclusions.* Using time-domain simulations, we have demonstrated that a spatially asymmetric resonant microcavity filled with a phase-change material (VO$_2$ in our case) acts as an optical diode (a nonlinear optical isolator) displaying unidirectional transmittance within a wide range of input light intensity. The same layered structure can also be seen as a highly asymmetric optical limiter with the forward limiting threshold different from the backward one by orders of magnitude. Above the respective (forward or backward) limiting threshold, the structure becomes highly reflective, which prevents overheating and significantly increases power-handling capabilities. The above features can be very desirable for infrared unidirectional valves and/or directional power limiters or switches [44-46].

*Acknowledgements*
We acknowledge support from the Office of Naval Research (MK: Grant No. N00014-16-1-2556; RT and TK: Grant No. N00014-16-1-2803) and Air Force Office of Scientific Research (IV: AFOSR 18RYCORO13).

Figures and Figure Captions

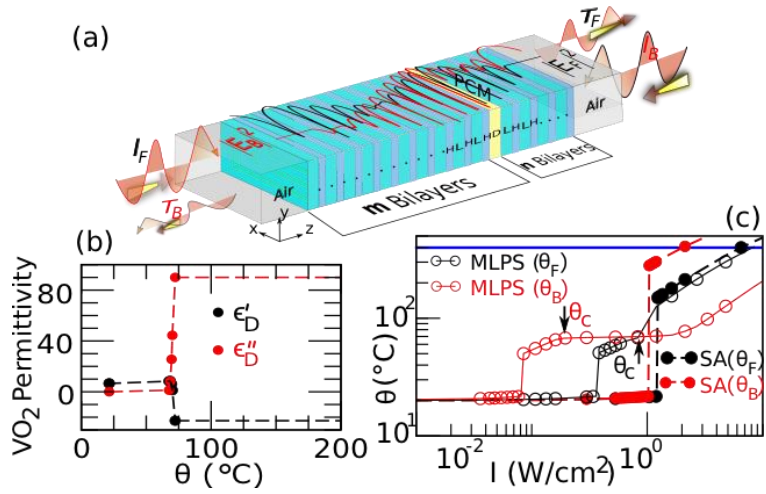

Fig. 1(a). The proposed multi-layered photonic structure (MLPS) consisting of quarter-wavelength alternating dielectric layers with an embedded phase-change material (PCM) defect layer, D, placed such that it breaks the mirror symmetry of the photonic structure. (b) The variation of ε' and ε" of the PCM, vanadium dioxide ($VO_2$), with temperature θ and $\lambda_0 = 10.5$ μm (the data for the temperature dependence of permittivity of $VO_2$ during the heating process are taken from Ref. [32]). (c) The calculated temperature increase at the center of the PCM layer in our MLPS versus incident irradiance, for forward (open black symbols) and backward (open red symbols) incident radiation, compared with that of a stand-alone (SA) PCM layer of the same thickness (solid red and black symbols). Note the value of the critical irradiance that triggers the phase-transition at θ is different for forward and backward incident radiation. In these simulations, the MLPS consists of $(LH)^6[VO_2](LH)^2$ quarter-wavelength layers with $L$ indicating a low-index dielectric layer (ZnS) and $H$ indicating a high index dielectric layer (Si). The incident wave has wavelength $\lambda_0 = 10.5$ μm. The SA layer (filled circles) consists of the same quarter wavelength $VO_2$ layer on a ZnS substrate. The associated transport characteristics of the SA structure are also shown for comparison purposes.

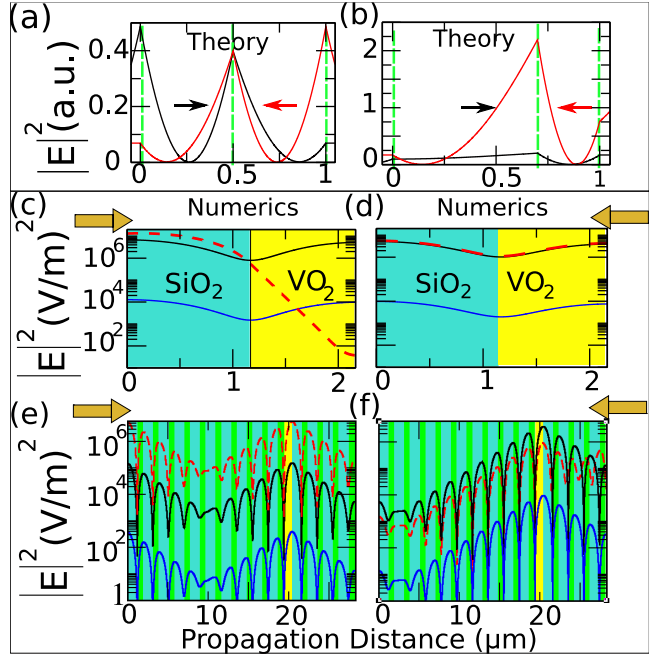

Fig. 2 (Top panel): Distribution of E-field intensity obtained analytically using Eqs. (1) and (2) for three (a) symmetrically and (b) asymmetrically positioned delta-function layers (vertical dashed green lines). The black and red curves denote the forward and backward directions of the incident wave, respectively. The x-axis indicates propagation distance in units of the cavity length $z_3 = 1$. (Bottom panel) Spatial electric-field intensity profiles in a stand-alone (SA) ZnS-$VO_2$ layer configuration are compared with the proposed MLPS (*m = 10 and n = 4*, bilayers) with an asymmetrically located defect layer (yellow bar) for the cases when a CW signal is incident in the forward (c & e), and backward (d & f) directions, respectively. The solid blue, black, and dashed red curves denote varying incident irradiances of 2.656e-5 (blue), 1.06e-2 (black), and 0.39 (red) W/$cm^2$.

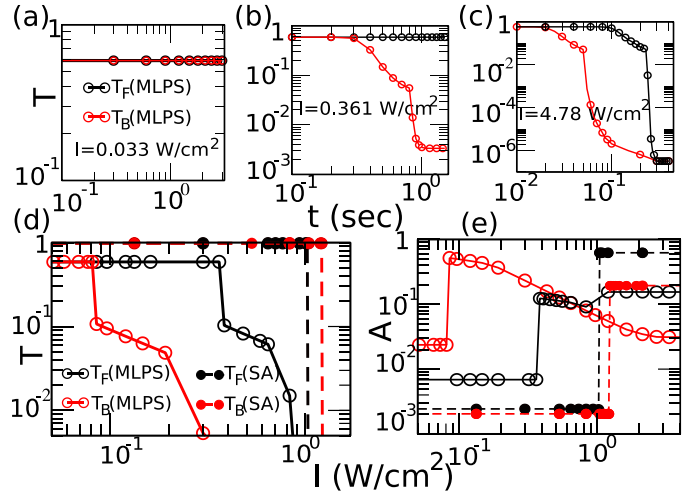

Fig 3. (a)-(c). Transient evolution of transmittance, $T(t)$, with increasing irradiances, $I=$ 0.033 (a), 0.361 (b) and 4.78 (c) W/cm$^2$ for the cases when light is incident in the forward (black symbols) and backward (red symbols) directions towards the MLPS (for m = 6, n = 2 bilayers) with asymmetrically located PCM defect layer. Note that for all the simulated results the background temperature is the ambient temperature, 293.15 K. (d)-(e) Transmittance and absorbance under the steady-state scenario with increasing irradiances in the case of MLPS, evaluated for cases when light enters the structure in the forward (F) and backward (B) directions. The dashed curve indicates the case of a single ZnS-VO$_2$ SA defect layered structure with the same defect thickness as the case of the MLPS. Note the reduction in $A$ with increasing incident $I$ in the case of MLPS when light is incident in the forward direction (i.e., from right).